\documentclass[twocolumn,superscriptaddress,amsmath,amssymb,fleqn,10pt]{wlscirep}
\usepackage{mathrsfs}
\usepackage{amssymb}
\usepackage{amsmath} 
\usepackage{graphicx}
\usepackage{amsfonts}
\usepackage{dcolumn}
\usepackage{array}
\usepackage{amsfonts}
\usepackage{threeparttable}
\usepackage{multirow}
\usepackage{threeparttable}
\usepackage{chngpage}
\usepackage{latexsym}

\title{A hybrid multiscale coarse-grained method for dynamics on complex networks}

\author[1,2,3,*]{Chuansheng Shen}
\author[4]{Hanshuang Chen}
\author[5]{Zhonghuai Hou}
\author[1,3,+]{J\"urgen Kurths}

\affil[1]{Department of Physics, Humboldt University, Berlin, 12489, Germany}
\affil[2]{Department of Physics, Anqing Normal University, Anqing, 246011, China}
\affil[3]{Potsdam Institute for Climate Impact Research, Potsdam, 14473, Germany}
\affil[4]{School of Physics and Material Science, Anhui University, Hefei, 230039, China}
\affil[5]{Hefei National Laboratory for Physical Sciences at Microscales, \& Department of Chemical Physics, University of Science and Technology of China, Hefei, 230026, China}

\affil[*]{chuansheng.shen@pik-potsdam.de}
\affil[+]{Juergen.Kurths@pik-potsdam.de}

\begin{abstract}
Brute-force simulations for dynamics on very large networks are quite expensive. While phenomenological treatments may capture some macroscopic properties, they often ignore important microscopic details. Fortunately, one may be only interested in the property of local part and not in the whole network.  Here, we
propose a hybrid multiscale coarse-grained(HMCG) method which combines a fine Monte
Carlo(MC)  simulation on the part of nodes of interest with  a more coarse Langevin dynamics  on the rest part.
We demonstrate the validity of our method by analyzing
the equilibrium Ising model and the nonequilibrium susceptible-infected-susceptible model. It is
found that HMCG not only works very well in reproducing the
phase transitions and critical phenomena of the microscopic models, but also accelerates the evaluation of
dynamics with significant computational savings compared to microscopic MC
simulations directly for the whole networks. The proposed method is general and can be applied to a
wide variety of networked systems just adopting
 appropriate microscopic simulation methods and coarse
graining approaches.
\end{abstract}

\begin{document}

\flushbottom
\maketitle
%
%
\thispagestyle{empty}

\section*{Introduction}

   Complex networks have been recently one of the most active research topics in
statistical physics and closely related
disciplines\cite{RMP02000047,AIP02001079,PRP06000175,PRP08000093,RMP08001275}.
The dynamics of networks and their topology are usually associated with
multiscale processes spanning from microscopic via mesoscopic, to
macroscopic level \cite{Nat2010098102,PNAS20096483,JCP2009245106}, 
like human multiscale mobility networks \cite{PNAS200921484}, module
networks \cite{PNAS2010099}, multilayer networks \cite{PRP2014122},
interconnected networks \cite{PNAS20148351}, and networks of
networks  \cite{NATPHY2012040}, etc. Although computer simulation
provides a powerful tool for studying and understanding complex
multiscale phenomena, brute-force simulations, such as Monte
Carlo(MC) simulation \cite{Lan2000}, and kinetic MC simulation
\cite{JPC1977812340}, are quite expensive and hence computationally
prohibited for simulating large networked systems. While
phenomenological models, such as mean-field description which need much less computational effort, may capture
certain properties of the system, but often ignore micro- and meso-scopic
details and fluctuation effects that may be important near critical
points. Therefore a promising way is to develop multiscale theory
and approaches, aiming at significantly accelerating the dynamical
evolution while properly preserving even microscopic information of
interest.

Recently, much efforts have been devoted to searching for coarse graining
(CG) approaches. Renormalization transformations have been used to
reduce the size of self-similar networks, while preserving the most
relevant topological properties of the original ones
\cite{PRL04016701,NTR05000392,PRL06018701,PRL08148701}. Gfeller and
Rios proposed a spectral technique to obtain a
CG-network which can reproduce the random walk and synchronization
dynamics of the original network \cite{PhysRevLett.99.038701,PRL08174104}. Kevrekidis
\emph{et al.} developed equation-free multiscale computational
methods to accelerate simulation using a coarse time-stepper
\cite{CMS03000715}, which has been successfully applied to study the
CG dynamics of oscillator networks \cite{PRL06144101}, gene
regulatory networks \cite{JCP06084106}, and adaptive epidemic
networks \cite{EPL08038004}. Very recently, we have proposed a
degree-based CG ($d$-CG) \cite{PRE10011107} approach and a
stength-based CG ($s$-CG) \cite{PhysRevE.83.066109} approach to
study the critical phenomena of the Ising model, the
susceptible-infected-susceptible (SIS) epidemic model and the  $q$-state Potts
model on complex networks.
 However, all of the works mentioned
above always coarse-grain the whole network. In fact, on the one
hand, most real-world networks are very very large
\cite{RMP06001213}. The higher the coarse graining, the more 
 information is lost. On the other hand, for specific purpose, we
often concern about the local dynamics of some nodes of interest and not about the entire nodes. However, the dynamics of a local part are certainly
influenced by that of the rest nodes of the network due to the
 interactions between connected individuals. Therefore
a natural question arise, could we simulate a part of interest at a fine level 
and  treat the rest one simultaneously at a CG level, while retaining the microscopic information of interest?

   To address the above question, in the present work, we develop
a hybrid multiscale  coarse-grained(HMCG) method to simulate phase transitions
of the networked Ising model and the
SIS model, which are often taken
as paradigms of equilibrium and non-equilibrium systems
respectively. First, according to the focus of interest, the network
is divided into two parts, where the part of interest nodes is
 named the \emph{core}, and the part of rest ones is called the \emph{periphery}. MC
simulations and Langevin equations (LE) are then performed on the \emph{core} and
the \emph{periphery}, respectively. Extensively numerical simulations show
that our HMCG method
works very well in reproducing the phase diagrams and fluctuations
of the microscopic models, while the LE does
not. Especially, our HMCG method accelerates the systems' dynamical
evolution much more than that of microscopic simulations.

\section*{Results}
Without loss of generality,  the underlying network is constructed as follows: starting from a random network with $N$ nodes and $N\langle k\rangle/2$  edges, where $\langle k\rangle$ is the average degree, then the network is split into two parts,  the \emph{core} consisting of $r_{core}N$ nodes,   and the \emph{periphery} with $(1-r_{core})N$ nodes, $r_{core}$ denotes 
the ratio of the number of nodes inside the \emph{core} to that of entire network.  We introduce the parameter $u$ as the density of the  inter-edges connecting the \emph{core} and the \emph{periphery},  and $p_c$ as the proportion of  edges inside the \emph{core} to the total number of intra-edges inside both of the \emph{core} and the \emph{periphery}.
We employ the HMCG method which combines a fine MC simulation with  a coarse Langevin dynamics as the fine level method and
 the CG method to treat the \emph{core} and the \emph{periphery} respectively (refer for details to the Methods section).

\subsection*{Application to the networked Ising model}

To evaluate the potential of the HMCG method, we begin with the networked Ising model, a typical example of an equilibrium system.
In a given network, each node is endowed with a spin
variable $s_i$ that can be either $+1$ (up) or $-1$ (down). The
Hamiltonian of the system is given by
\begin{equation}
H =  - J\sum\limits_{i < j} {A_{ij} s_i s_j }  - h\sum\limits_i {s_i
},{\kern 6pt}  (s =  \pm 1,i,j = 1, \cdots ,N) 
\label{eq:eq1}
\end{equation}
where $J$ is the coupling constant and $h$ is the external magnetic
field. The elements of the adjacency matrix of the network take $A_{ij} =
1$ if nodes $i$ and $j$ are connected and $A_{ij} =0$ otherwise. The
degree, that is the number of neighboring nodes, of node $i$ is
defined as $ k_i = \sum\nolimits_{j = 1}^N {A_{ij} }.$

\begin{figure}[h!]
\centering
\includegraphics[width=\linewidth]{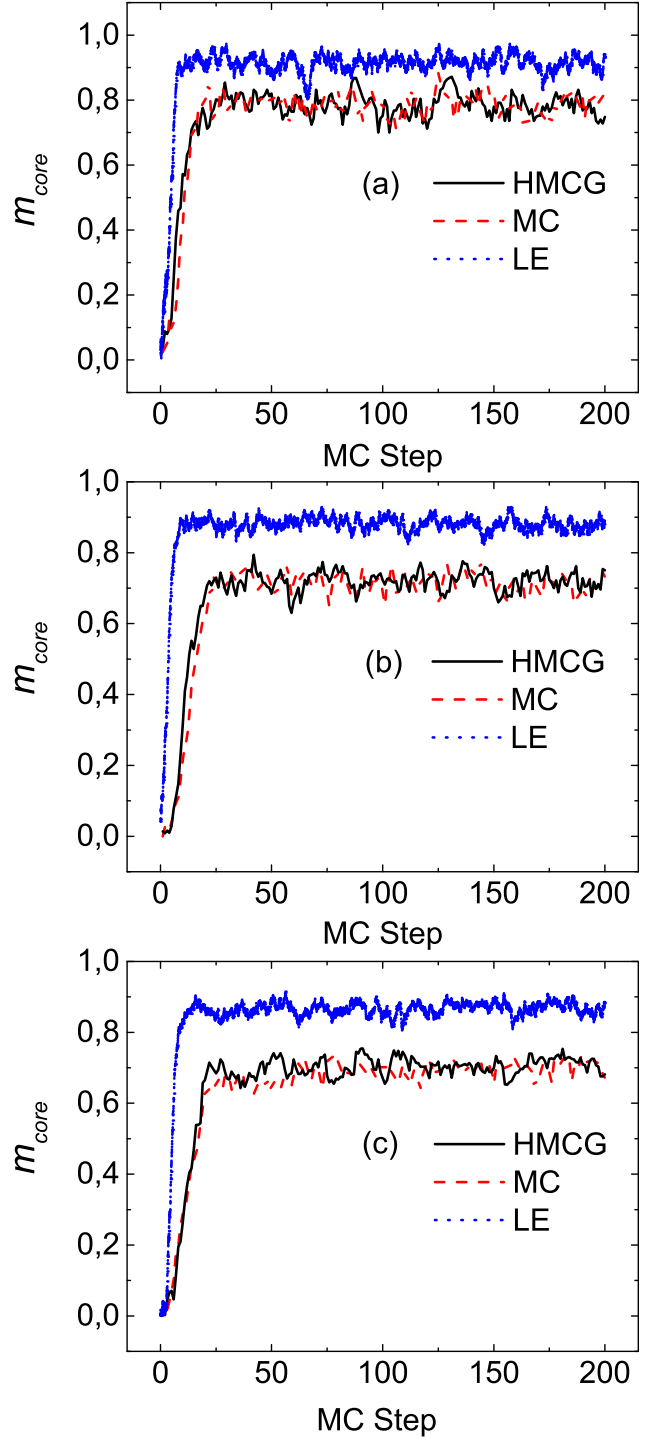}
\caption{Typical time evolutions of the magnetization $m_{core}$ in Ising model at $T$=2.5 (in unit of $J/k_B$) and $h$=0 for (a)  $r_{core}$=0.05, (b)  $r_{core}$=0.1 and (c)  $r_{core}$=0.15, where solid lines, dashed and dotted lines indicated HMCG method, MC simulations and LE approach. Other
parameters are $N$=10,000, $\langle k\rangle $=6, $u$=0.01, $p_c$=0.6. }
\label{fig:fig1}
\end{figure}

MC simulations with Glauber dynamics and LE are
performed on the \emph{core} and the \emph{periphery} respectively (see methods for the details). Generally, with
increasing the temperature $T$, the system undergoes a second-order
phase transition at the critical value $T_c$ from an ordered
state to a disordered one.  Figure \ref{fig:fig1} plots typical time evolutions of the magnetization
$m_{core}=\sum\nolimits_{i\in \mathscr{C}}{s_i/(r_{core}N)}$ in (1) for different size $r_{core}$ at
$T$=2.5 (in unit of $J/k_B$) and $h$=0.
For both HMCG and the microscopic MC simulations, the systems attain the steady
states associated with fluctuating noise after transient time. It is
clear that they are in good agreement in the steady-state values of
$m_{core}$, as well as their fluctuating amplitudes for both simulations
cases at different size $r_{core}$, while the LE is not.

\begin{figure}[h]
\centering
\includegraphics[width=\linewidth]{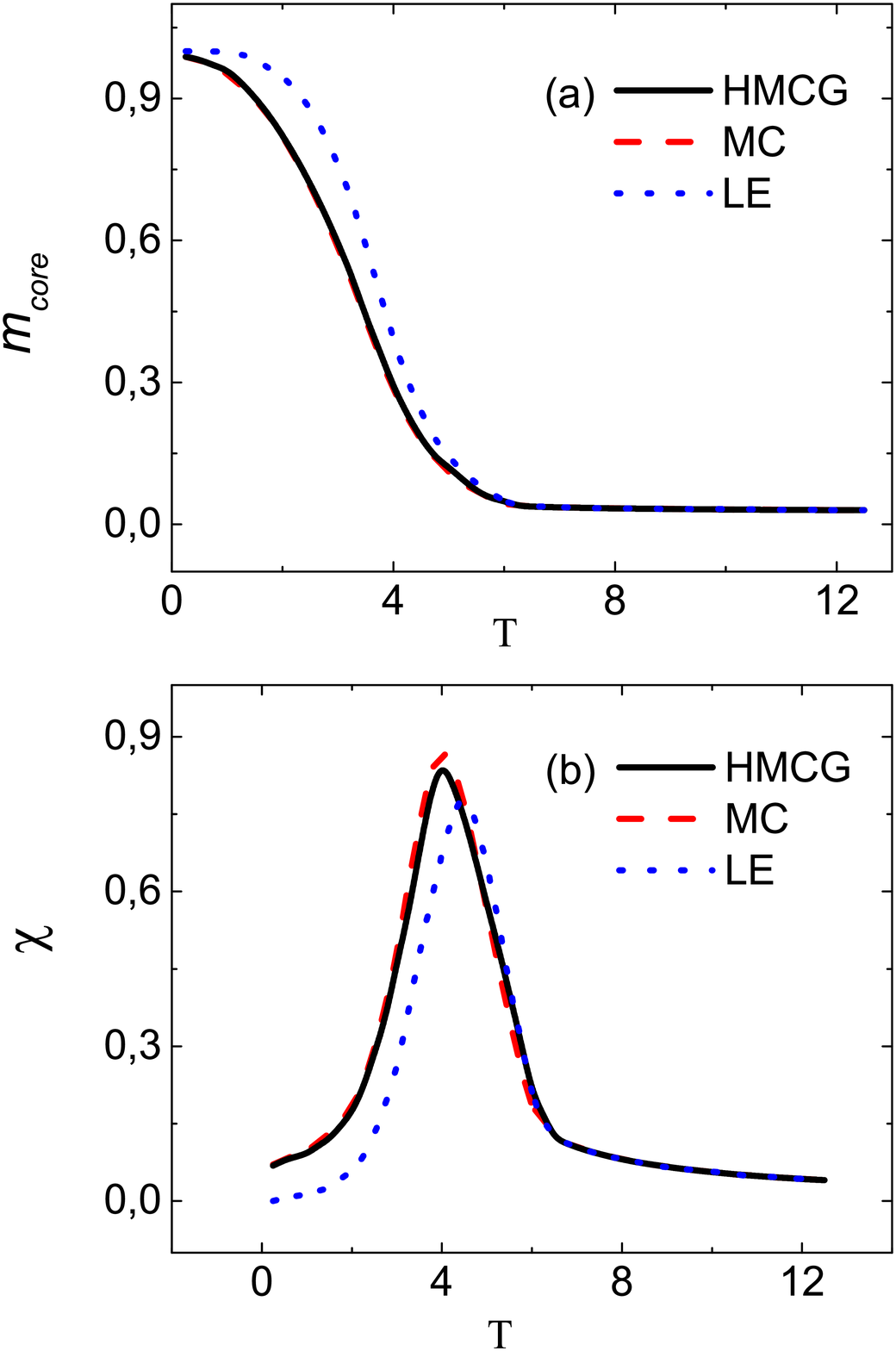}
\caption{$m_{core}$ and $\chi$ as functions of $T$ for the Ising model on
complex networks. The solid lines, dashed and dotted lines correspond
to the HMCG, MC and LE simulation results,
respectively. Other parameters are the same as in Figure \ref{fig:fig1}.}
\label{fig:fig2}
\end{figure}

Furthermore, $m_{core}$  as a function of $T$ is plotted  in Figure \ref{fig:fig2}(a), obtained from our HMCG
method, micro-MC simulations and LE. Again, the
agreements between HMCG and MC are excellent, further demonstrating
the validity of HMCG method. In order to ensure that the microscopic
configurations are nearly identical between both methods, we
calculate the susceptibility $ \chi  = r_{core}N (\langle m_{core}^2 \rangle -
\langle m_{core}\rangle ^2)/(k_BT)  $, since  $ \chi $ is related to the variance of the
magnetization according to the fluctuation-dissipation
theorem, and compare $ \chi$ as a function of $T$ in
Figures \ref{fig:fig2}(b). Very good agreement is again seen between HMCG
and MC method.

\subsection*{Application to the networked SIS model}

Concerning nonequilibrium scenarios, a prototype example is the
spreading dynamics of SIS models
\cite{May1992,Dal1999,RevModPhys.87.925} on
complex network as mentioned above, where
individuals inside each node run stochastic infection dynamics as
follows:
\begin{equation}\label{eqModel}
S + I\xrightarrow{\lambda }2I,{\kern 10pt} I\xrightarrow{\mu}S
\end{equation}
The first reaction indicates that each susceptible (S) individual with the state variable $\sigma=0$
becomes infected upon encountering one infected (I) individual with $\sigma=1$ at a
rate $\lambda$. The second one reflects that the
infected individuals are cured and become again susceptible at a
rate $\mu$. For simplicity (yet without loss of generality), we set $\mu=1$.
 In this model, a significant and general result is that the system
undergoes an absorbing-to-active phase transition at a critical value
$\lambda_c$ with an increasing infectious rate $\lambda$.

\begin{figure}[h]
\centering
\includegraphics[width=\linewidth]{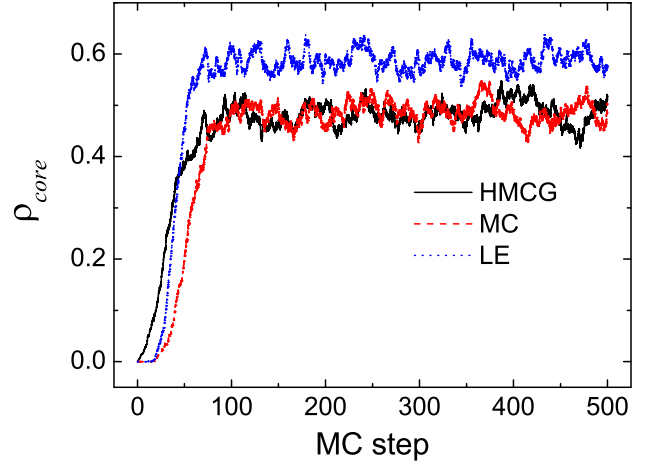}
\caption{Typical time evolutions $\rho_{core}$ of the density of infected
nodes inside the \emph{core} in SIS model at $\lambda=0.8$ for HMCG method, microscopic MC
dynamics, and LE. Other parameters are $N$=10,000, $\langle k \rangle $=6,
$r_{core}$=0.1, $u$=0.01, $p_c$=0.6. }
\label{fig:fig3} 
\end{figure}

 Our numerical simulation starts from
a random configuration with several nodes being infected. After an
initial transient regime, the system will evolve into a steady state
with a constant average density of infected nodes. The steady
density of infected nodes $\rho$ is computed by averaging over at
least 50 different initial configurations and at least 20
different network realizations for a given $\lambda$.
Figure \ref{fig:fig3} compares typical time evolutions $\rho_{core}=\sum\nolimits_{i\in \mathscr{C}}
{\sigma_i}/(r_{core}N)$ of the
density of infected nodes inside the \emph{core}  at $\lambda=0.8$,
 for the HMCG method, microscopic MC
dynamics, and Langevin approach indicated by the solid, dashed and
dotted line respectively. Excellent agreement between HMCG
and MC is shown.

\begin{figure}[h]
\centering
\includegraphics[width=\linewidth]{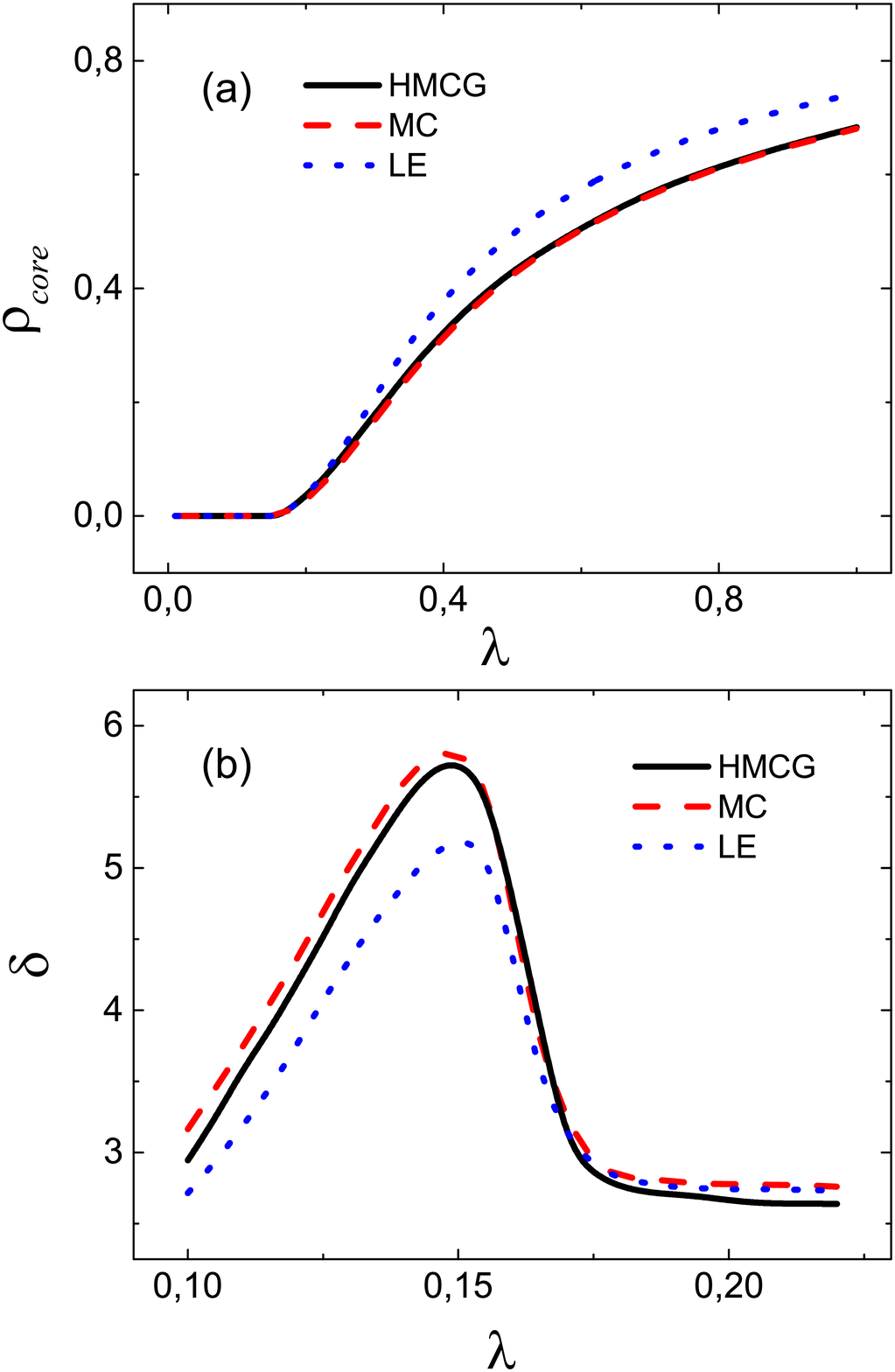}
\caption{$\rho_{core}$ and $\delta$ as functions of $T$ for SIS model on
complex networks. The solid, dashed and dotted lines correspond to
the results of HMCG, MC and LE, respectively. Other
parameters are same as in Figure \ref{fig:fig3}. }
\label{fig:fig4}
\end{figure}

To further validate the effect of our method, we compare the calculated results of
$\rho_{core}$ and normalized susceptibility $ \delta =r_{core}N(\langle \rho_{core} ^2 \rangle -
\langle \rho_{core} \rangle ^2 )/\langle \rho_{core} \rangle $  as a
function of $\lambda$  in Figures \ref{fig:fig4} (a) and (b) respectively, obtained by the HMCG method, the microscopic MC
dynamics and LE. Clearly, the agreement between the
HMCG and the microscopic MC results remains excellent, while the
LE fails. On the one hand, as shown in Fig.
\ref{fig:fig4} (a), the HMCG can reproduce well the main
characteristic: the system undergoes a phase transition at a certain
threshold rate $\lambda_c$, above which $\rho_{core}$ monotonically
increases from zero indicating the epidemic spreading, otherwise,
i.e., $\lambda<\lambda_c$ , the system stays in a healthy state
with $\rho_{core}=0$. On the other hand, both HMCG and MC methods exhibit a
maximum susceptibility $\delta$ at the threshold  $\lambda_c$,  as can be seen in Fig.
\ref{fig:fig4} (b), which
suggests that the microscopic configurations of the HMCG method are nearly
identical to those of the original model. Note that the normalized susceptibility
$\delta$ adopted here is different from the traditional definition $
\delta =r_{core}N (\langle \rho_{core} ^2 \rangle - \langle \rho_{core} \rangle ^2 )
$\cite{Dic1999}, because it leads to clearer numerical
results, while preserving all the scaling properties of the usual
definition \cite{PhysRevE.86.041125}.

\section*{Discussion}

\begin{figure}[h]
\centering
\includegraphics[width=\linewidth]{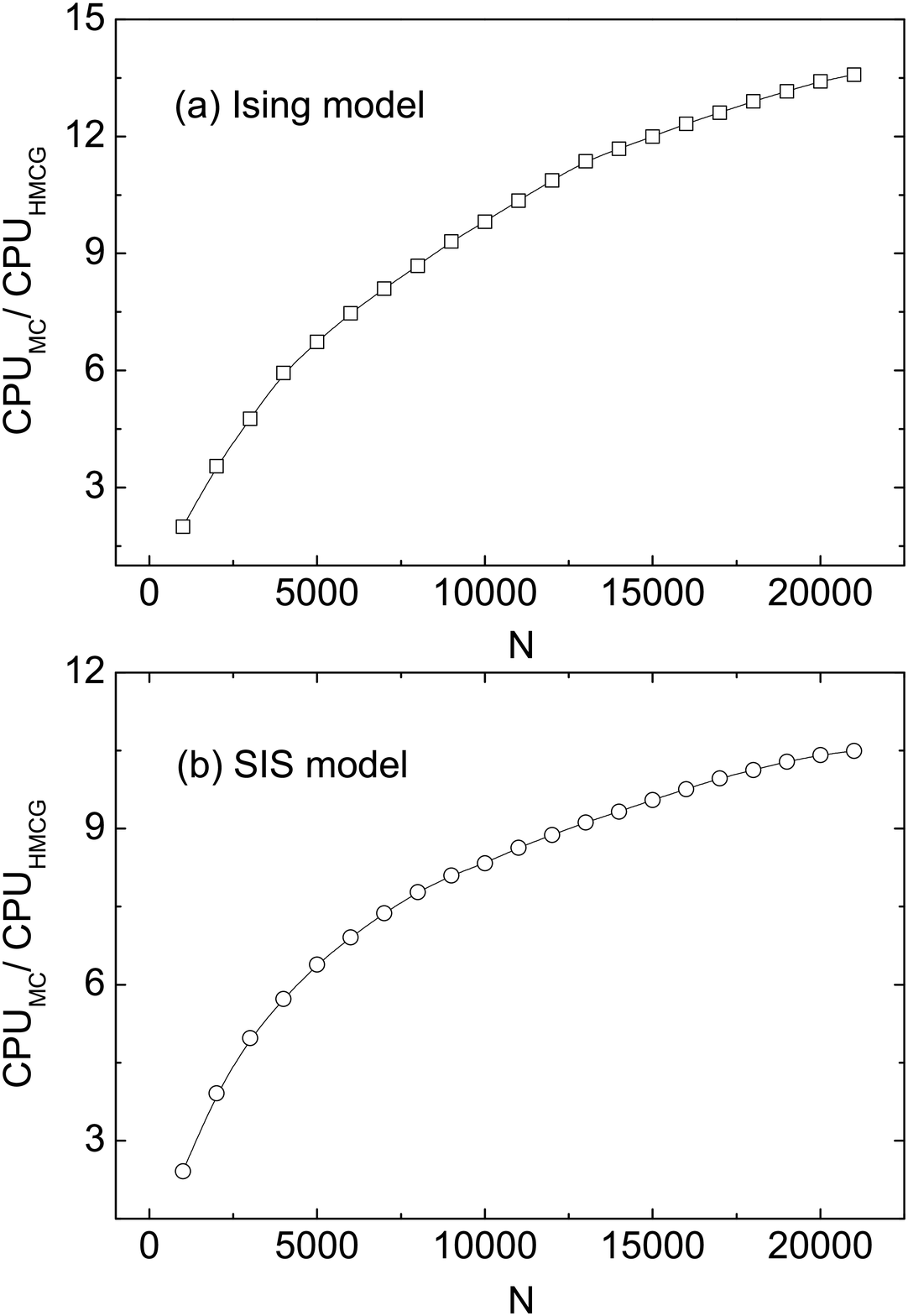}
\caption{The proportion $CPU_{MC}/CPU_{HMCG} $ as a function of $N$ for the Ising model (a) and for the SIS model (b), where $CPU_{MC}$ and $CPU_{HMCG} $ denote the CPU time resulted from
MC and HMCG method respectively. Other parameters in (a) and (b) are same as in
Figure \ref{fig:fig1} and Figure \ref{fig:fig3} respectively.}
\label{fig:fig5}
\end{figure}

Note that the main goal to develop the multiscale coarse grained method is to  improve the computational efficiency. We count the CPU time resulted from microscopic MC simulations
and from the HMCG method, indicated by $CPU_{MC}$ and $CPU_{HMCG} $
respectively, and compare them in Figures \ref{fig:fig5}  (a) for the Ising model and (b) for the SIS model. 
It can be seen
that, on the one hand, the HMCG method provides substantial computational savings compared to the microscopic MC simulations for
the same size of network. On the other hand, 
the ratio $CPU_{MC}/CPU_{HMCG} $ shows an apparently monotonic dependence on $N$, suggesting that for a given size of the \emph{core}, the larger the network becomes,
the larger computational savings are. One may approximately estimate the total
savings by $  \frac{{CPU_{MC} }}{{CPU_{HMCG} }} \approx \frac{{N
\times \langle k\rangle /2}}{{r_{core}N \times \langle k_{core}\rangle /2 + t_L
}} $, where $ \langle k_{core}\rangle   = p_c (1 - u)\langle k\rangle $, denoting the average
 degree of the group of interest, and $t_L $ denotes the computational
 cost of LE for the rest group. Generally, $t_L\ll r_{core}N  \times \langle k_{core}\rangle  /2
 $, thus $t_L $ can be neglected and $ \frac{{CPU_{MC} }}{{CPU_{HMCG} }}
\approx \frac{{1 }}{{r_{core} \times p_c(1-u) }} $ is obtained. Specifically, for $ N = 10,000$,  $\langle k\rangle = 6$,  $r_{core} = 0.1$, $u=0.01$ and $p_c=0.6$, we obtain
$CPU_{MC}/CPU_{HMCG} \approx 16.8 $. Obviously, the computational savings are mainly dependent on the relative size of the interest part compared with
that of the entire, and on the density of links of intra-\emph{core} and inter-parts. 
 Therefore, if the original network is far larger than the part of interest, the efficiency of our method will become more significant.

In this study, a hybrid multiscale coarse-grained method is proposed that combines a
fine simulations for the part of interest with a CG level for the
rest of network. Specifically, microscopic MC simulations 
and LE are employed to treat
both parts respectively. Extensively numerical simulations
demonstrate that both the networked Ising model and SIS model, two
paradigms for equilibrium and nonequilibrium systems, show a very
good agreement of the HMCG and MC method. By comparing CPU times for HMCG
and MC method, we find that a large computational cost is
saved. 
The HMCG method can not only be suitable to random
networks, scale-free networks without or with strength correlation,
but also to dense networks and sparse networks
\cite{PhysRevLett.113.208702}. The proposed method thus is general,
very easy to implement, and directly related to the microscopics
models. Therefore, this method can be applied to a wide variety of
networked systems just choosing appropriate microscopic simulation
methods, such as kinetic MC method, molecular dynamics, and
 other CG approaches instead of MC method and
LE respectively in view of different real-world scenarios.

\section*{Methods}

\begin{figure}[h]
\centering
\includegraphics[width=\linewidth]{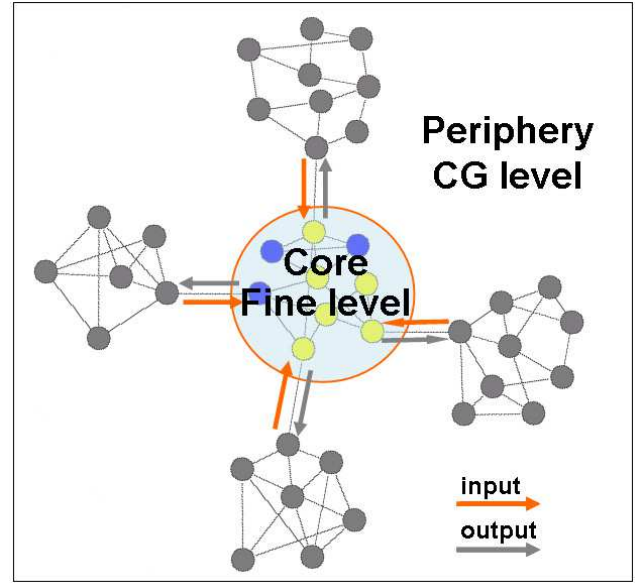}
\caption{Schematic illustration of the hybrid
multiscale coarse-grained method.  The original network is divided into two parts,
i.e., the part of module of interest named the \emph{core}, and the part of
rest ones named the \emph{periphery}, which are treated at a fine level and
a CG level respectively.} 
\label{fig:fig6} 
\end{figure}

To account for the idea and procedure of the HMCG method, we give a  
schematic illustration by a module network consisting of five connected random
subgraphs with different topologies, as shown in Figure \ref{fig:fig6}.
 The main idea is as follows: to
capture the local information and achieve high efficiency in the
simulation, the network is divided into two parts, i.e., the \emph{core} which is the module of interest and the \emph{periphery} which consists of the rest ones.  Then a fine
level simulation and a CG level one are performed on the part of interest and the other part of rest respectively. Here, we adopt a
microscopic simulation of detailed allowed by classical MC dynamics
and a LE to treat the two parts respectively.

The main steps are summarized below:

 (i) \emph{Identifying the network parts}. According to the requirement of interest, the network
is split into two parts, i.e., the \emph{core}, and
the rest one the \emph{periphery}.  We then employ $\mathscr{C}$,
$\mathscr{P}$ and $\mathscr{B}$ to denote the adjacency matrices of
intra-\emph{core}, intra-\emph{periphery}, and inter-parts
respectively. Note that $\mathscr{P}$ and $\mathscr{B}$ are coarse grained, while $\mathscr{C}$ preserves so as to to pay close attention to the part of the original system which is different from other CG methods.

 (ii) \emph{Determining the input and output}. In view of the \emph{core} is the part of interest, we define the  flux from the \emph{periphery} to the
\emph{core} as  the input, where the flux is the product of the mean-field of
 the \emph{periphery} and the average links between the two parts, and the
 output is the reverse process.

 (iii) \emph{Performing simulations}. Simulation methods such as MC dynamics,
 kinetic MC dynamics,  molecular dynamics, etc, are performed on the \emph{core} and the \emph{periphery}
 with a more coarse method, e.g., the LE, spectrum coarse graining, $d$-CG, $s$-CG and other CG methods. Specifically, here, MC
 simulation and LE are employed as the fine level method and
 the CG method to treat the part of interest and the rest one respectively. 

(iv) \emph{Improving the method}. The CPU time
of the HM method and that of microscopic MC simulations are counted and compared as well as the accuracy of the results, and then the method is improved by optimizing the algorithm.

\subsection*{HMCG method for networked Ising model}
The MC simulation at the microscopic level follows standard Glauber
dynamics: At each step, we randomly selected a node from the group
of interest nodes and try to flip its spin with an acceptance
probability $1/(1+exp(\Delta E/(k_BT)))$, where $\Delta E$ is the associated
change of energy due to the flipping process, $k_B$ the
Boltzmann constant and $T$ the temperature. 

A simple recipe of the Glauber algorithm is described as follows:

(1) Choose an initial state

(2) Choose a node $i$ at random, $ i \in \mathscr{C}$

(3) Calculate the energy change $\Delta E=\Delta E_{in} + \Delta E_{out}$,
resulting from the part of interest and
the rest one respectively supposed the spin of node $i$ is flipped.
Since $ \mathscr{C}$ is known for the part of interest, $\Delta E_{in}$ can be calculated directly by the
microscopic simulations, while $\Delta E_{out}$ should be estimated through
the mean field coupling between the spin of node $i$ and the net
magnetization $m'$(to be derived in the next step) of the rest part because that $ \mathscr{P}$ is coarse grained

(4) Generate a random number $r$ such that $0 < r < 1$

(5) If $r <1/(1+exp((\Delta E)/(k_BT)))$, flip
the spin of node $i$

(6) Go to (2)

Next, we will derive the fluctuation-driven LE for
$m'$. The average change of magnetization $m'$ due to spin-flipping
can be written as follows
\begin{equation}
\langle d m'\rangle  = d m' _ \uparrow   \times p_ \uparrow \times
W_{ \uparrow , \downarrow }  + d m'_ \downarrow \times p_ \downarrow
\times W_{ \downarrow , \uparrow }
\end{equation}
where $ d m'_ \uparrow   =  - \frac{2}{(1-r_{core})N}$  denotes the net change
of magnetization if a up-spin turns to down-spin, and $d m'_
\downarrow = \frac{2}{(1-r_{core})N}$ denotes the reverse process. $p_
\uparrow = \frac{{1 + m'}}{2}$ and $ p_ \downarrow = \frac{{1 -
m'}}{2} $ represent the probabilities of up spins and down spins
respectively. $ W_{ \uparrow , \downarrow }$ and $W_{ \downarrow ,
\uparrow } $ represent the transition probabilities from up-spin to
down-spin and its reverse process respectively. According to the
rule of Glauber dynamics, they take the forms
\begin{subequations}
\begin{eqnarray}
 W_{ \uparrow , \downarrow }  &=& \frac{1}{{1 + e^{\Delta E'/T} }} =
\frac{1}{2}(1 - \tanh (\frac{{\Delta E'}}{{2T}})) \\
 W_{
\downarrow , \uparrow } &=&  \frac{1}{{1 + e^{-\Delta E'/T} }} =
\frac{1}{2}(1 + \tanh (\frac{{\Delta E'}}{{2T}}))
 \end{eqnarray}
\end{subequations}
where $ \Delta E'= 2\left( {
u m' + \sum\nolimits_{i \in \mathscr{C} } \mathscr{C} s_i
} \right)/(1-r_{core})N $ is the energy change due to flipping a up-spin
within the rest group. Therefore, Eq.(3) can be rewritten as
\begin{equation}
\begin{array}{l}
 \langle d m' \rangle = \frac{1}{(1-r_{core})N}( - m' + \tanh (\frac{{\Delta E'}}{{2T}}))
 \end{array}
\end{equation}

Then, we calculate the mean square deviation of $m'$
\begin{equation}
\begin{array}{l}
 \langle d m'^2 \rangle  = \frac{4}{{(1-r_{core})^2N^2 }}  \times \frac{{1 + m'}}{2}
  \times \frac{1}{2}(1 - \tanh (\frac{{\Delta E'}}{{2T}})) \\
  {\kern 35pt} + \frac{4}{{(1-r_{core})^2N^2 }}
   \times \frac{{1 - m'}}{2} \times \frac{1}{2}(1 + \tanh (\frac{{\Delta E'}}{{2T}})) \\
 {\kern 30pt}  = \frac{1}{{(1-r_{core})^2N^2 }}(2 - 2m'\tanh (\frac{{\Delta E'}}{{2T}})) \\
 \end{array}
\end{equation}
When we adopt $ d t = 1 /(1-r_{core})N$, the fluctuation-driven Langevin
 equation can be obtained
\begin{equation}
\begin{array}{l}
\frac{{d m'}}{{d t}} =  - m' + \tanh (\frac{{\Delta E'}}{{2T}})  \\
   {\kern 28pt} +
\sqrt {\frac{1}{{(1-r_{core})N }}(2 - 2m'\tanh (\frac{{\Delta E'}}{{2T}}))}
\xi (t)
 \end{array}
\end{equation}
where $ \xi (t)$ is a Gaussian white-noise satisfying $\langle \xi
(t) \rangle=0$ and $\langle \xi (t)\xi (t')\rangle =\delta(t-t')$.

\subsection*{HMCG method for networked SIS model}

To begin, the subgraph of interest is treated with the microscopic
MC dynamics as follows

(1) Choose an initial state

(2) Randomly choose a node $i$, $ i \in \mathscr{C}$

(3) If $i$ is susceptible, calculate the total number of infected
individuals $nI$ of its nearest neighbors, which contains within and
without the part of interest, denoted by $nI_{in}$ and $nI_{out}$
respectively. Notice that $nI_{in}=\sum\nolimits_{j \in
\mathscr{C} }\mathscr{C}_{ij}\sigma_j $ can be calculated directly by the microscopic
simulation, while $nI_{out}=\sum\nolimits_{j \in \mathscr{P}
}\mathscr{P}_{i,j}\sigma_j$ is estimated through the mean field coupling with the
average density of infected nodes $\rho'$ inside the rest part, since $\mathscr{P}$ is coarse grained. If $i$
is infectious, go to (6)

(4) Generate a random number $r_1$ such that $0 < r_1 < 1$

(5) If $r_1 <\lambda nI dt$, $i$ is infected, then go to (2)

(6) Generate a random number $r_2$ such that $0 < r_2 < 1$

(7) If $r_2 <dt$, $i$ becomes susceptible, then go to (2)

Then, we will derive the fluctuation-driven LE of
$\rho'$ for the rest subgraph. Following
Ref.\cite{PhysRevE.79.036110}, one has
\begin{equation}
\begin{array}{l}
 \frac{{d\rho '}}{{dt}} =  - \rho ' + \lambda (1 - \rho ')\left( {\langle k_p\rangle \rho ' + \sum\nolimits_{i \in \mathscr{C} } \mathscr{C}\sigma_i } \right) \\
   {\kern 24pt} + \sqrt {\frac{1}{{(1-r_{core})N }}[\rho ' + \lambda (1 - \rho ')(\langle k_p\rangle \rho ' + \sum\nolimits_{i \in \mathscr{C} } \mathscr{C}\sigma_i  )]} \xi (t) \\
 \end{array}
\end{equation}
where $\langle k_p\rangle=(1-u)(1-r_{core}p_c)\langle k \rangle /(1-r_{core}) $ denotes the average degree of the
subgraph of the rest, $\sigma_i$ is the state variable of node
$i$, $\sigma_i=0, 1$ represent susceptible and infectious respectively.
$\xi (t)$ is also a Gaussian white-noise satisfying $\langle \xi (t)
\rangle=0$ and $\langle \xi (t)\xi (t')\rangle =\delta(t-t')$.


\begin{thebibliography}{10}
\expandafter\ifx\csname url\endcsname\relax
  \def\url#1{\texttt{#1}}\fi
\expandafter\ifx\csname urlprefix\endcsname\relax\def\urlprefix{URL }\fi
\providecommand{\bibinfo}[2]{#2}
\providecommand{\eprint}[2][]{\url{#2}}

\bibitem{RMP02000047}
\bibinfo{author}{Albert, R.} \& \bibinfo{author}{Barab\'{a}si, A.-L.}
\newblock \bibinfo{title}{Statistical mechanics of complex networks}.
\newblock \emph{\bibinfo{journal}{Rev. Mod. Phys.}}
  \textbf{\bibinfo{volume}{74}}, \bibinfo{pages}{47} (\bibinfo{year}{2002}).

\bibitem{AIP02001079}
\bibinfo{author}{Dorogovtsev, S.~N.} \& \bibinfo{author}{Mendes, J. F.~F.}
\newblock \bibinfo{title}{Evolution of networks}.
\newblock \emph{\bibinfo{journal}{Adv. Phys.}} \textbf{\bibinfo{volume}{51}},
  \bibinfo{pages}{1079--1139} (\bibinfo{year}{2002}).

\bibitem{PRP06000175}
\bibinfo{author}{Boccaletti, S.}, \bibinfo{author}{Latora, V.},
  \bibinfo{author}{Moreno, Y.}, \bibinfo{author}{Chavez, M.} \&
  \bibinfo{author}{Hwang, D.-U.}
\newblock \bibinfo{title}{Complex networks: Structure and dynamics}.
\newblock \emph{\bibinfo{journal}{Phys. Rep.}} \textbf{\bibinfo{volume}{424}},
  \bibinfo{pages}{175--308} (\bibinfo{year}{2006}).

\bibitem{PRP08000093}
\bibinfo{author}{Arenas, A.}, \bibinfo{author}{D\'iaz-Guilera, A.},
  \bibinfo{author}{Kurths, J.}, \bibinfo{author}{Moreno, Y.} \&
  \bibinfo{author}{Zhou, C.}
\newblock \bibinfo{title}{Synchronization in complex networks}.
\newblock \emph{\bibinfo{journal}{Phys. Rep.}} \textbf{\bibinfo{volume}{469}},
  \bibinfo{pages}{93--153} (\bibinfo{year}{2008}).

\bibitem{RMP08001275}
\bibinfo{author}{Dorogovtsev, S.~N.}, \bibinfo{author}{Goltsev, A.~V.} \&
  \bibinfo{author}{Mendes, J. F.~F.}
\newblock \bibinfo{title}{Critical phenomena in complex netwokrs}.
\newblock \emph{\bibinfo{journal}{Rev. Mod. Phys.}}
  \textbf{\bibinfo{volume}{80}}, \bibinfo{pages}{1275} (\bibinfo{year}{2008}).

\bibitem{Nat2010098102}
\bibinfo{author}{Ahn, Y.-Y.}, \bibinfo{author}{Bagrow, J.~P.} \&
  \bibinfo{author}{Lehmann, S.}
\newblock \bibinfo{title}{Link communities reveal multiscale complexity in
  networks}.
\newblock \emph{\bibinfo{journal}{Nature}} \textbf{\bibinfo{volume}{466}},
  \bibinfo{pages}{761--764} (\bibinfo{year}{2010}).

\bibitem{PNAS20096483}
\bibinfo{author}{Serrano, M.~A.}, \bibinfo{author}{Bogun\'a, M.} \&
  \bibinfo{author}{Vespignani, A.}
\newblock \bibinfo{title}{Extracting the multiscale weighted networks}.
\newblock \emph{\bibinfo{journal}{Proc. Natl. Acad. Sci. USA}}
  \textbf{\bibinfo{volume}{106}}, \bibinfo{pages}{6483--6488}
  (\bibinfo{year}{2009}).

\bibitem{JCP2009245106}
\bibinfo{author}{Jang, H.}, \bibinfo{author}{Na, S.} \& \bibinfo{author}{Eom,
  K.}
\newblock \bibinfo{title}{Multiscale network model for large protein dynamics}.
\newblock \emph{\bibinfo{journal}{J. Chem. Phys.}}
  \textbf{\bibinfo{volume}{131}}, \bibinfo{pages}{245106}
  (\bibinfo{year}{2009}).

\bibitem{PNAS200921484}
\bibinfo{author}{Balcan, D.} \emph{et~al.}
\newblock \bibinfo{title}{Multiscale mobility networks and the spatial
  spreading of infectious diseases}.
\newblock \emph{\bibinfo{journal}{Proc. Natl. Acad. Sci. USA}}
  \textbf{\bibinfo{volume}{106}}, \bibinfo{pages}{21484--21489}
  (\bibinfo{year}{2009}).

\bibitem{PNAS2010099}
\bibinfo{author}{Girvan, M.} \& \bibinfo{author}{Newman, M. E.~J.}
\newblock \bibinfo{title}{Community structure in social and biological
  networks}.
\newblock \emph{\bibinfo{journal}{Proc. Natl. Acad. Sci. USA}}
  \textbf{\bibinfo{volume}{99}}, \bibinfo{pages}{761--764}
  (\bibinfo{year}{2010}).

\bibitem{PRP2014122}
\bibinfo{author}{Boccaletti, S.} \emph{et~al.}
\newblock \bibinfo{title}{The structure and dynamics of multilayer networks}.
\newblock \emph{\bibinfo{journal}{Phys. Rep.}} \textbf{\bibinfo{volume}{544}},
  \bibinfo{pages}{1--122} (\bibinfo{year}{2014}).

\bibitem{PNAS20148351}
\bibinfo{author}{Domenico, M.~D.}, \bibinfo{author}{Sol\'e-Ribalta, A.},
  \bibinfo{author}{G\'omez, S.} \& \bibinfo{author}{Arenas, A.}
\newblock \bibinfo{title}{Navigability of interconnected networks under random
  failures}.
\newblock \emph{\bibinfo{journal}{Proc. Natl. Acad. Sci. USA}}
  \textbf{\bibinfo{volume}{111}}, \bibinfo{pages}{8351--8356}
  (\bibinfo{year}{2014}).

\bibitem{NATPHY2012040}
\bibinfo{author}{Gao, J.}, \bibinfo{author}{Buldyrev, S.~V.},
  \bibinfo{author}{Stanley, H.~E.} \& \bibinfo{author}{Havlin, S.}
\newblock \bibinfo{title}{Networks formed from interdependent networks}.
\newblock \emph{\bibinfo{journal}{Nature phys.}} \textbf{\bibinfo{volume}{8}},
  \bibinfo{pages}{40--48} (\bibinfo{year}{2012}).

\bibitem{Lan2000}
\bibinfo{author}{Landau, D.~P.} \& \bibinfo{author}{Binder, K.}
\newblock \emph{\bibinfo{title}{A Guide to Monte Carlo Simulations in
  Statistcal Physics}} (\bibinfo{publisher}{Cambridge University Press},
  \bibinfo{address}{Cambridge}, \bibinfo{year}{2000}).

\bibitem{JPC1977812340}
\bibinfo{author}{Gillespie, D.~T.}
\newblock \bibinfo{title}{Exact stochastic simulation of coupled chemical
  reactions}.
\newblock \emph{\bibinfo{journal}{J. Phys. Chem.}}
  \textbf{\bibinfo{volume}{81}}, \bibinfo{pages}{2340--2361}
  (\bibinfo{year}{1977}).

\bibitem{PRL04016701}
\bibinfo{author}{Kim, B.~J.}
\newblock \bibinfo{title}{Geographical coarse graining of complex networks}.
\newblock \emph{\bibinfo{journal}{Phys. Rev. Lett.}}
  \textbf{\bibinfo{volume}{93}}, \bibinfo{pages}{168701}
  (\bibinfo{year}{2004}).

\bibitem{NTR05000392}
\bibinfo{author}{Song, C.}, \bibinfo{author}{Havlin, S.} \&
  \bibinfo{author}{Makse, H.~A.}
\newblock \bibinfo{title}{Self-similarity of complex networks}.
\newblock \emph{\bibinfo{journal}{Nature}} \textbf{\bibinfo{volume}{433}},
  \bibinfo{pages}{392} (\bibinfo{year}{2005}).

\bibitem{PRL06018701}
\bibinfo{author}{Goh, K.-I.}, \bibinfo{author}{G.Salvi},
  \bibinfo{author}{Kahng, B.} \& \bibinfo{author}{Kim, D.}
\newblock \bibinfo{title}{Skeleton and fractal scaling in complex networks}.
\newblock \emph{\bibinfo{journal}{Phys. Rev. Lett.}}
  \textbf{\bibinfo{volume}{96}}, \bibinfo{pages}{018701}
  (\bibinfo{year}{2006}).

\bibitem{PRL08148701}
\bibinfo{author}{Radicchi, F.}, \bibinfo{author}{Ramasco, J.~J.},
  \bibinfo{author}{Barrat, A.} \& \bibinfo{author}{S.Fortunato}.
\newblock \bibinfo{title}{Complex networks renormalization: Flows and fixed
  points}.
\newblock \emph{\bibinfo{journal}{Phys. Rev. Lett.}}
  \textbf{\bibinfo{volume}{101}}, \bibinfo{pages}{148701}
  (\bibinfo{year}{2008}).

\bibitem{PhysRevLett.99.038701}
\bibinfo{author}{Gfeller, D.} \& \bibinfo{author}{De~Los~Rios, P.}
\newblock \bibinfo{title}{Spectral coarse graining of complex networks}.
\newblock \emph{\bibinfo{journal}{Phys. Rev. Lett.}}
  \textbf{\bibinfo{volume}{99}}, \bibinfo{pages}{038701}
  (\bibinfo{year}{2007}).

\bibitem{PRL08174104}
\bibinfo{author}{Gfeller, D.} \& \bibinfo{author}{Rios, P. D.~L.}
\newblock \bibinfo{title}{Spectral coarse graining and synchronization in
  oscillator networks}.
\newblock \emph{\bibinfo{journal}{Phys. Rev. Lett.}}
  \textbf{\bibinfo{volume}{100}}, \bibinfo{pages}{174104}
  (\bibinfo{year}{2008}).

\bibitem{CMS03000715}
\bibinfo{author}{Kevrekidis, I.~G.} \emph{et~al.}
\newblock \bibinfo{title}{Equation-free, coarse-grained multiscale computation:
  Enabling microscopic simulators to perform system-level analysis}.
\newblock \emph{\bibinfo{journal}{Comm. Math. Sci.}}
  \textbf{\bibinfo{volume}{1}}, \bibinfo{pages}{715--762}
  (\bibinfo{year}{2003}).

\bibitem{PRL06144101}
\bibinfo{author}{Moon, S.~J.}, \bibinfo{author}{Ghanem, R.} \&
  \bibinfo{author}{Kevrekidis, I.~G.}
\newblock \bibinfo{title}{Coarse graining the dynamics of coupled oscillators}.
\newblock \emph{\bibinfo{journal}{Phys. Rev. Lett.}}
  \textbf{\bibinfo{volume}{96}}, \bibinfo{pages}{144101}
  (\bibinfo{year}{2006}).

\bibitem{JCP06084106}
\bibinfo{author}{Erbana, R.}, \bibinfo{author}{Kevrekidis, I.~G.},
  \bibinfo{author}{Adalsteinsson, D.} \& \bibinfo{author}{Elston, T.~C.}
\newblock \bibinfo{title}{Gene regulatory networks: A coarse-grained,
  equation-free approach to multiscale computation}.
\newblock \emph{\bibinfo{journal}{J. Chem. Phys.}}
  \textbf{\bibinfo{volume}{124}}, \bibinfo{pages}{084106}
  (\bibinfo{year}{2006}).

\bibitem{EPL08038004}
\bibinfo{author}{Gross, T.} \& \bibinfo{author}{Kevrekidis, I.~G.}
\newblock \bibinfo{title}{Robust oscillations in sis epidemics on adaptive
  networks: Coarse graining by automated moment closure}.
\newblock \emph{\bibinfo{journal}{EPL}} \textbf{\bibinfo{volume}{82}},
  \bibinfo{pages}{38004} (\bibinfo{year}{2008}).

\bibitem{PRE10011107}
\bibinfo{author}{Chen, H.~S.}, \bibinfo{author}{Hou, Z.~H.},
  \bibinfo{author}{Xin, H.~W.} \& \bibinfo{author}{Yan, Y.~J.}
\newblock \bibinfo{title}{Statistically consistent coarse-grained simulations
  for critical phenomena in complex networks}.
\newblock \emph{\bibinfo{journal}{Phys. Rev. E}} \textbf{\bibinfo{volume}{82}},
  \bibinfo{pages}{011107} (\bibinfo{year}{2010}).

\bibitem{PhysRevE.83.066109}
\bibinfo{author}{Shen, C.~S.}, \bibinfo{author}{Chen, H.~S.},
  \bibinfo{author}{Hou, Z.~H.} \& \bibinfo{author}{Xin, H.~W.}
\newblock \bibinfo{title}{Coarse-grained monte carlo simulations of the phase
  transition of the potts model on weighted networks}.
\newblock \emph{\bibinfo{journal}{Phys. Rev. E}} \textbf{\bibinfo{volume}{83}},
  \bibinfo{pages}{066109} (\bibinfo{year}{2011}).

\bibitem{RMP06001213}
\bibinfo{author}{Rabinovich, M.~I.}, \bibinfo{author}{Varona, P.},
  \bibinfo{author}{Selverston, A.~I.} \& \bibinfo{author}{Abarbanel, H. D.~I.}
\newblock \bibinfo{title}{Dynamical principles in neuroscience}.
\newblock \emph{\bibinfo{journal}{Rev. Mod. Phys.}}
  \textbf{\bibinfo{volume}{78}}, \bibinfo{pages}{1213--1264}
  (\bibinfo{year}{2006}).

\bibitem{May1992}
\bibinfo{author}{Anderson, R.} \& \bibinfo{author}{R.M.May}.
\newblock \emph{\bibinfo{title}{Infectious Diseases in Humans}}
  (\bibinfo{publisher}{Oxford University Press}, \bibinfo{address}{Oxford},
  \bibinfo{year}{1992}).

\bibitem{Dal1999}
\bibinfo{author}{Daley, D.~J.} \& \bibinfo{author}{Gani, J.}
\newblock \emph{\bibinfo{title}{Epidemic Modelling}}
  (\bibinfo{publisher}{Cambridge University Press},
  \bibinfo{address}{Cambridge}, \bibinfo{year}{1999}).

\bibitem{RevModPhys.87.925}
\bibinfo{author}{Pastor-Satorras, R.}, \bibinfo{author}{Castellano, C.},
  \bibinfo{author}{Van~Mieghem, P.} \& \bibinfo{author}{Vespignani, A.}
\newblock \bibinfo{title}{Epidemic processes in complex networks}.
\newblock \emph{\bibinfo{journal}{Rev. Mod. Phys.}}
  \textbf{\bibinfo{volume}{87}}, \bibinfo{pages}{925--979}
  (\bibinfo{year}{2015}).

\bibitem{Dic1999}
\bibinfo{author}{Marro, J.} \& \bibinfo{author}{Dickman, R.}
\newblock \emph{\bibinfo{title}{Nonequilibrium Phase Transitions in Lattice
  Models}} (\bibinfo{publisher}{Cambridge University Press},
  \bibinfo{address}{Cambridge}, \bibinfo{year}{1999}).

\bibitem{PhysRevE.86.041125}
\bibinfo{author}{Ferreira, S.~C.}, \bibinfo{author}{Castellano, C.} \&
  \bibinfo{author}{Pastor-Satorras, R.}
\newblock \bibinfo{title}{Epidemic thresholds of the
  susceptible-infected-susceptible model on networks: A comparison of numerical
  and theoretical results}.
\newblock \emph{\bibinfo{journal}{Phys. Rev. E}} \textbf{\bibinfo{volume}{86}},
  \bibinfo{pages}{041125} (\bibinfo{year}{2012}).

\bibitem{PhysRevLett.113.208702}
\bibinfo{author}{Karrer, B.}, \bibinfo{author}{Newman, M. E.~J.} \&
  \bibinfo{author}{Zdeborov\'a, L.}
\newblock \bibinfo{title}{Percolation on sparse networks}.
\newblock \emph{\bibinfo{journal}{Phys. Rev. Lett.}}
  \textbf{\bibinfo{volume}{113}}, \bibinfo{pages}{208702}
  (\bibinfo{year}{2014}).

\bibitem{PhysRevE.79.036110}
\bibinfo{author}{Bogu\~n\'a, M.}, \bibinfo{author}{Castellano, C.} \&
  \bibinfo{author}{Pastor-Satorras, R.}
\newblock \bibinfo{title}{Langevin approach for the dynamics of the contact
  process on annealed scale-free networks}.
\newblock \emph{\bibinfo{journal}{Phys. Rev. E}} \textbf{\bibinfo{volume}{79}},
  \bibinfo{pages}{036110} (\bibinfo{year}{2009}).

\end{thebibliography}

\section*{Acknowledgements}

This work was supported by the National Basic
Research Program of China (2013CB834606) and by the National Natural
Science Foundation of China (Grants No. 11475003, 11205002
and No. 21473165). C.S. was also funded by the China Scholarship Council (CSC) and 
Natural Science Foundation of Anhui Province (Grant No.1408085MA09).

\section*{Author contributions}

C.S. and H.C. conceived the study, constructed the hybrid multiscale coarse-grained method and performed the numerics. C.S., H. C., Z.H. and J.K. discussed the results, drew conclusions, prepared, edited and reviewed the manuscript.

\section*{Additional information}
Competing financial interests: The authors declare no competing financial interests.

\end{document}